\documentclass[conference]{IEEEtran}
\IEEEoverridecommandlockouts

\usepackage{cite}
\usepackage{amsmath,amssymb,amsfonts}
\usepackage{algorithmic}
\usepackage{graphicx}
\usepackage{braket}
\usepackage{textcomp}
\usepackage{url}
\usepackage{hyperref}
\usepackage{xcolor}
\usepackage{comment}
\usepackage{balance}

\def\BibTeX{{\rm B\kern-.05em{\sc i\kern-.025em b}\kern-.08em
    T\kern-.1667em\lower.7ex\hbox{E}\kern-.125emX}}
\begin{document}

\title{Quantum-assisted Gaussian process regression using random Fourier features\\

\thanks{We want to gratefully acknowledge funding from the Research Council of Finland, Project No. 350221.}
}

\author{\IEEEauthorblockN{Cristian A. Galvis-Florez}
\IEEEauthorblockA{\textit{Department of Electrical}\\
\textit{Engineering and Automation,} \\
\textit{Aalto University}\\
Espoo, Finland \\
cristian.galvis@aalto.fi}
\and
\IEEEauthorblockN{Ahmad Farooq}
\IEEEauthorblockA{\textit{Quantum Algorithms and Software,} \\
\textit{VTT Technical Research Centre of Finland}\\
Espoo, Finland \\
ahmad.farooq@vtt.fi}
\and
\IEEEauthorblockN{Simo Särkkä}
\IEEEauthorblockA{\textit{Department of Electrical}\\
\textit{Engineering and Automation,} \\
\textit{Aalto University}\\
Espoo, Finland \\
simo.sarkka@aalto.fi}
}

\maketitle

\begin{abstract}
Probabilistic machine learning models are distinguished by their ability to integrate prior knowledge of noise statistics, smoothness parameters, and training data uncertainty. A common approach involves modeling data with Gaussian processes; however, their computational complexity quickly becomes intractable as the training dataset grows. To address this limitation, we introduce a quantum-assisted algorithm for sparse Gaussian process regression based on the random Fourier feature kernel approximation. We start by encoding the data matrix into a quantum state using a multi-controlled unitary operation, which encodes the classical representation of the random Fourier features matrix used for kernel approximation. We then employ a quantum principal component analysis along with a quantum phase estimation technique to extract the spectral decomposition of the kernel matrix. We apply a conditional rotation operator to the ancillary qubit based on the eigenvalue. We then use Hadamard and swap tests to compute the mean and variance of the posterior Gaussian distribution. We achieve a polynomial-order computational speedup relative to the classical method.
\end{abstract}


\section{Introduction}
Fields such as robotics, geophysics, data mining, and machine learning make extensive use of Gaussian Processes (GPs) \cite{DFR:13:IEEE_TPAMI,SSH:13:IEEE_SPM}. Multidimensional nonlinear functions can be modeled using probabilistic, non-parametric Gaussian Process regression methods characterized by their mean and covariance functions \cite{RW:06:MIT}. The dependence structure between function values at different input points is described by the covariance function. However, computing the covariance of the posterior distribution for $N$ datapoints is computationally challenging, with its complexity and memory requirements scaling as $O(N^3)$ and $O(N^2)$, respectively. This complexity arises due to the inversion of the kernel matrix $N \times N$. To address this issue, promising approaches such as inducing points, the Hilbert-space approximation, and random Fourier features (RFF) have been introduced to approximate the kernel matrix of the stationary function \cite{GCRV:10:JMLR,SS:20:SC,CR:05:JMLR,RR:07:ANIPS}. Stationary covariant kernels are attractive for modeling spatial processes because they possess desirable properties \cite{RW:06:MIT}. They exhibit isotropy, meaning that the covariance depends only on the distance between the points and not their direction.  This approximation transforms the kernel matrix into a low-rank approximation of size $N \times M$, where $M \ll N$. This reduced-rank approximation of the kernel function reduces the computational complexity to $O(NM^2)$ or $O(M^3)$ (for likelihood and prediction, respectively) \cite{RR:07:ANIPS, SS:20:SC}.

Recently, quantum computers have been used to accelerate classical machine learning tasks \cite{BWPRWL:17:Nat,RML:14:PRL,P:18:QCNISQ}. The first attempt to speed up the Gaussian process regression using a quantum computer was introduced in \cite{ZFF:19:PRA}. This algorithm, built upon the quantum matrix inversion algorithm known as the HHL algorithm, is designed specifically for $s$-sparse matrices with a condition number denoted as $\kappa$. It achieves a desired level of accuracy represented by $\epsilon$ and exhibits a run-time that scales roughly as $O(\log(N) \kappa^2 s^2 / \epsilon)$. This algorithm assumes that the data are already uploaded as a quantum state and that the quantum unitaries can be implemented efficiently. Later, the authors in \cite{CYGYLGL:22:PRA} addressed these caveats and proposed a quantum algorithm for Gaussian process regression that maintains reasonable computational complexity. Recently, the authors in \cite{FGS:24:PRA} proposed quantum-assisted Gaussian process regression using Hilbert space approximation, achieving a polynomial speedup over the classical algorithm. This technique can be extended to compute the numerical integration of analytical intractable functions \cite{GFS:25:TMLR}. 
 
In this paper, we develop a quantum algorithm for low-rank approximation using RFF to accelerate the computational task.  Our approach begins with the classical preparation of a $N \times M$ data matrix denoted as $X$. We encode this data matrix into a quantum state by employing Hadamard gates and multi-controlled rotation unitaries. We then utilize the analysis of the principal components of quantum data \cite{LMR:14:NPh}, which allows us to extract the dominant eigenvectors and eigenvalues from a quantum register. We use conditional rotations based on the eigenvalue register before applying Hadamard and SWAP tests to derive the mean and variance of reduced-rank Gaussian process regression \cite{SSP:16:PRA}. Our approach is similar to the Hilbert space method, but instead of evaluating the eigenfunctions along with its eigenvalues, we took random samples from the spectral density, which is the Fourier transform of the stationary covariance function.

We can prepare the quantum state in two ways: first, by preparing the random Fourier feature basis classically and then uploading the basis using the standard quantum state preparation method or amplitude approximation method. Second, random sampling from the spectral density of the covariance function can be done classically,
and then a multi-controlled unitary rotation can be used to upload the sinusoidal and cosine functions. We follow the second approach in this paper.
In \cite{FGS:24:PRA}, preparing the quantum state for the dataset requires a computational measurement, and preparing the quantum state requires a probabilistic algorithm. The quantum state preparation of the random Fourier feature provided here is a deterministic method.
We also provide numerical simulations to demonstrate the effectiveness of our algorithm. 
The contribution of our paper is to reduce the computational complexity from $O(NM^2)$ to $O(NM \log(M) \epsilon^{-3} \kappa^2)$, similar to that of \cite{FGS:24:PRA} with the deterministic quantum state preparation method. This represents a polynomial speedup compared to the classical algorithm.

The structure of the paper is as follows: first, we provide the Gaussian process regression and how its kernel function can be approximated using random Fourier features in Sec. \ref{Sec:2}. Next, we present our detailed method for using quantum computing in Gaussian process regression in Sec \ref{sec:3}. Sec \ref{Sec:4} discusses the computational complexity of our method and also provides numerical simulations to demonstrate the effectiveness of our algorithm. Finally, we conclude our discussion in Sec \ref{Sec:5}.

\section{\label{Sec:2} Kernel approximation of Gaussian Process regression using random Fourier feature} 

In this section, we will first provide an overview of Gaussian process regression. We then provide details on how the kernel function can be approximated using random Fourier features.

\subsection{Gaussian Process Regression}
 Gaussian process regression is a non-parametric probabilistic regression technique used for modeling and predicting data. Given a set of observed data points $\mathcal{D} = {(\mathbf{x}_i, y_i)}_{i=1}^N$ containing $d$-dimensional inputs $\{\textbf{x}_{i}\}_{i=1}^{N}$ and corresponding outputs $\{y_{i}\}_{i=1}^{N}$, we assume that the model functions $f$ are realizations of a Gaussian random process prior \cite{RW:06:MIT}. The output observations are also corrupted by Gaussian noise $\varepsilon_i \sim \mathcal{N}\left(0,\sigma_{n}^2\right)$
\begin{eqnarray}
   f \sim \mathcal{GP}\left(0, k\left(\textbf{x}_i,\textbf{x}_j\right) \right),\\
   y_{i}=f\left(\mathbf{x}_{i}\right)+\varepsilon_i,
\end{eqnarray}
where $k\left(\textbf{x}_i,\textbf{x}_j\right) $ is the covariance matrix (kernel) defined by a positive semi-definite matrix $k: \Omega \times \Omega \rightarrow \mathbb{R}$. We can select our own choice of the kernel function. One commonly used stationary kernel function is the Gaussian radial basis function (RBF) kernel, also known as the squared exponential kernel with length scale $l$ defined as \cite{PTS:04:KCB} 
\begin{equation}
k\left(\textbf{x}_{i},\textbf{x}_{j}\right) = \sigma_{f}^{2}\exp\left(-\frac{1}{2l^2}||\textbf{x}_i - \textbf{x}_j||^2\right).
\end{equation}
where $\sigma_{f}$ and $l$ are the hyper-parameters of the kernel function. This leads to the kernel matrix $\mathbf{K}\in \mathrm{R}^{N \times N}$ with entries $K_{ij}=k\left(\textbf{x}_{i},\textbf{x}_{j}\right)$.

GP regression aims to predict the mean and variance of the posterior distribution given new input data points $\mathbf{x}_{*}$.  The mean prediction $E\left[f_*\right]$ and its variance $\mathrm{V}\left[f_*\right] $ also Gaussian distribution can be computed as follows
\begin{equation}
    p\left(f_* \mid \mathbf{x}_{*}\right) \sim \mathcal{N}\left(E\left[f_*\right], \mathrm{V}\left[f_*\right] \right).
\end{equation}
The expressions for the mean prediction and variance in GP regression are as follows

\begin{eqnarray}
    	E\left[f_*\right] &=& \mathbf{k}_*^T\left(\mathbf{K}+\sigma_n^2 I\right)^{-1} \mathbf{y},\label{eq: Classic GPR} \\
	\mathrm{V}\left[f_*\right] &=& k\left(\mathbf{x}_{*}, \mathbf{x}_{*}\right)-\mathbf{k}_*^T\left(\mathbf{K}+\sigma_n^2 I\right)^{-1} \mathbf{k}_* .
\end{eqnarray}
where $\mathbf{k}_*=k\left(\textbf{x}_{*},\textbf{x}_{i}\right)$ is a vector of covariance values between the new input data points $\mathbf{x}_{*}$ and $i$th training data points $\mathbf{x}_{i}$. 
\subsection{\label{sec:level2}Kernel function approximation}
We can approximate the stationary kernel function with low-rank approximation using random Fourier feature \cite{RR:07:ANIPS, HDS:18:JMLR}. The stationary property implies that the covariance function depends exclusively on the difference between data points, denoted as $\tau = \textbf{x}_i - \textbf{x}_j$.  The stationary kernel function $k\left(\tau\right)$ can be expressed as the Fourier transform of a positive measure using Bochner's theorem \cite{STE:99:book}, which allows us to write
\begin{equation}
     k(\tau) = \frac{1}{(2\pi)^d}\int_{\mathbb{R}^d} S(\omega)e^{i\omega^\top \tau}d\omega, \label{eq: K Bochner}
\end{equation}
being $S(\omega)$ the spectral density of the kernel.

This states that stationary kernels can be represented as the Fourier transform of a positive measure, where the power spectrum of the positive measure is proportional to a probability measure. By evaluating the proportionality constant at $\tau = 0$, we have:
\begin{equation}
S\left(s\right)=k\left(0\right)p_{S}\left(s\right)=\sigma_{0}^{2}p_{S}\left(s\right),
\end{equation}
where $S\left(s\right)$ is proportional to a multivariate probability density in $s$. We can rewrite the stationary covariance function as an expectation \cite{GCRV:10:JMLR}

\begin{equation}
k\left(\tau\right)=k(\textbf{x}_i,\textbf{x}_j) = \sigma_0^2 \mathbb{E}_{p_s}\left[e^{2\pi \dot{\iota} s^T \textbf{x}_i}\left(e^{2\pi \dot{\iota} s^T \textbf{x}_j}\right)^*\right], \label{spectrum}
\end{equation}
where $p_{S}\left(s\right)$ represents the distribution over frequencies $s$. Eq.~(\ref{spectrum}) provides an exact expansion of the covariance function as an expectation of a product of complex exponential with respect to the distribution $p_{S}\left(s\right)$ over their frequencies.
To approximate this expression, we can employ a Monte Carlo simulation using a finite set of frequencies. Since the power spectrum is symmetric around zero, we can sample valid frequency pairs as $\left(s_r, -s_r\right)$. This approach preserves the exact expansion of Eq.~(\ref{spectrum}), as the imaginary terms cancel out
 \begin{eqnarray}
     k\left(\textbf{x}_{i},\textbf{x}_{j}\right)\simeq\frac{\sigma_{0}^{2}}{2M}  \sum_{r=1}^{M}\left[e^{2 \pi \dot{\iota} s_{r}^{T}\textbf{x}_{i}}\left(e^{2 \pi \dot{\iota} s_{r}^{T}\textbf{x}_{j}} \right)^{*}\right. \nonumber\\\left.+\left(e^{2 \pi \dot{\iota} s_{r}^{T}\textbf{x}_{i}} \right)^{*} e^{2 \pi \dot{\iota} s_{r}^{T}\textbf{x}_{j}}\right].
 \end{eqnarray}
Note that the previous approximation corresponds to a Monte Carlo integration of Eq. \eqref{eq: K Bochner}, where the samples are taken from the spectral density $S(\omega)$.
 
Simplifying further, we obtain
\begin{equation}
k\left(\textbf{x}_i,\textbf{x}_j\right) \simeq \frac{\sigma_{0}^{2}}{M}  \sum_{r=1}^{M} \cos\left(2\pi s_{r}^{T}\left(\textbf{x}_{i}-\textbf{x}_{j}\right)\right),
\end{equation}
where $s_r$ represents spectral frequencies drawn from $p_{S}\left(s\right)$. This allows us to express the covariance function in the following form
\begin{align}
   k\left(\mathbf{x}_{i},\mathbf{x}_{j}\right)&=\frac{\sigma_{0}^{2}}{M}  \sum_{r=1}^{M} \cos\left(2\pi s_{r}^{T}\left(\mathbf{x}_{i}-\mathbf{x}_{j}\right)\right)\notag\\
   &=\frac{\sigma_{0}^{2}}{M} \phi^{T}\left(\mathbf{x}_{i}\right) \phi\left(\mathbf{x}_{j}\right),
\end{align}
where $\phi(\mathbf{x})$ is a column vector defined as
\begin{eqnarray} \label{Random Fourier Feature}
      \phi\left(\mathbf{x}\right)=\left(\cos\left(2\pi s_{1}^{T} \mathbf{x}\right), \sin\left(2\pi s_{1}^{T} \mathbf{x}\right), \right. \nonumber\\ \left.\ldots,  \cos\left(2\pi s_{M}^{T} \mathbf{x}\right),  \sin\left(2\pi s_{M}^{T} \mathbf{x}\right)\right).
\end{eqnarray}

Now, we can make a data matrix $X \in \mathrm{R}^{N\times 2M}$ in the following way
\begin{equation} \label{data Matrix}
   \mathbf{X} =\left[ \phi\left(\textbf{x}_{1}\right), \phi\left(\textbf{x}_{2}\right), \ldots, \phi\left(\textbf{x}_{N}\right)\right].
\end{equation}
By representing the kernel function in this trigonometric basis form, we can approximate the original kernel using $M$ inducing points, making the computations more tractable while maintaining reasonable accuracy. The expression for evaluating the mean and variance of Gaussian process regression  becomes
\begin{eqnarray}
          E\left[f_{*}\right]=\phi_{*}^{T}\left(\mathbf{X}^{T}\mathbf{X}+
        \sigma_{n}^{2} \mathrm{I}\right)^{-1}\mathbf{X}^{T}  y\\
         \mathrm{V}\left[f_*\right] =\sigma_{0}^{2} \phi_{*}^{T}\left(\mathbf{X}^{T}\mathbf{X}+\sigma_{n}^{2} \mathrm{I} \right)^{-1} \phi_{*},
\end{eqnarray}
where $\phi_{*}$ is the vector of random Fourier features corresponding to the new input data points $\mathbf{x}_*$. Singular value decomposition of the mean and variance can be written as
\begin{eqnarray}
    \label{Mean}
     E\left[f_{*}\right]=\sum_{r=1}^{R} \frac{\lambda_{r}}{\lambda_{r}^{2}+\sigma_{n}^{2}} \mathbf{X}_{*}^{T}\mathbf{V}_{r} \mathbf{U}_{r}^{\top}\mathbf{y}. \nonumber\\
         \mathrm{V}\left[f_*\right]=\sigma_{0}^{2}\sum_{r=1}^{R} \frac{1}{\lambda_{r}^{2}+\sigma_{n}^{2}} \mathbf{X}_{*}^{T}\mathbf{V}_{r} \mathbf{V}_{r}^{\top}\mathbf{X}_{*}.
\end{eqnarray} 
Here $\Sigma\in R^{R\times R}$ is a diagonal matrix containing the real singular values $\lambda_1,\lambda_2, \ldots,\lambda_R$ and $U\in R^{N \times R}$ (and $V\in R^{R \times 2M}$) are the left (right) orthogonal matrices with columns corresponding to the singular values $\lambda_{R}$. 

\section{\label{sec:3} Method}

This section introduces the quantum-assisted Gaussian process regression model for sparse datasets using an RFF kernel approximation. Our approach includes several key subroutines, including quantum state preparation, quantum principal component analysis, quantum phase estimation algorithm, conditional rotational unitary operations, and the Hadamard and SWAP test. 

\subsection{Quantum State}
 A pure quantum state is mathematically defined by a unit vector and represented as a ket vector. The $n$-qubit pure quantum state can be defined as
\begin{equation}
    \ket{\psi}=\sum_{i=1}^{n}\alpha_{i}\ket{i},
\end{equation}
where $\sum_{i=1}^{n}|\alpha_{i}|^{2}=1$ and $\ket{i}$ are orthonormal computational basis $\{\ket{0\cdots 0}=\ket{0},\cdots,\ket{1\cdots 1}=\ket{2^{n}-1}\}$ . The mixed quantum state known as the density matrix is defined as the mixture of the pure states occurring with different probabilities as
\begin{equation}
    \rho=\sum_{i}p_{i}\ket{\psi_{i}}\bra{\psi_{i}}.
\end{equation}
where $\sum_{i}p_{i}=1$.
\subsection{Encoding dataset into quantum computer}

We start by vectorizing a matrix into a single-column vector to encode classical data into a quantum state.  The encoded quantum state is given by

    \begin{equation} \label{Encoded state}
   \ket{ \psi_{\mathbf{X}}}=\frac{1}{2^{NM}}\begin{bmatrix}
      \cos\left(2\pi s_{1}^{T} \textbf{x}_{1}\right)\\ \sin\left(2\pi s_{1}^{T} \textbf{x}_{1}\right)\\\vdots \\ \sin\left(2\pi s_{M}^{T} \textbf{x}_{1}\right)  \\ \cos\left(2\pi s_{M}^{T} \textbf{x}_{2}\right)\\  \vdots \\\cos\left(2\pi s_{M}^{T} \textbf{x}_{N}\right)\\ \sin\left(2\pi s_{M}^{T} \textbf{x}_{N}\right)
   \end{bmatrix}.
\end{equation}
To generate the desired quantum state $\ket{X}$ on a superconducting quantum computer, we employ Hadamard gates denoted as $H=\frac{1}{\sqrt{2}}\begin{bmatrix}
    1& 1 \\1 &-1
\end{bmatrix}$, and conditional multicontrolled  $R_{y}$ gates denoted as  $R_{y}\left(\theta\right)=\begin{bmatrix}
    \cos\left(\theta\right)& -\sin\left(\theta\right) \\\sin\left(\theta\right)  &\cos\left(\theta\right)
\end{bmatrix}$ condition on all the possible combinations of the qubits register. The initial state of the entire quantum system is
\begin{equation}
    \ket{\phi}_{1}=\ket{0\cdots 0}.
\end{equation}
Following this, the Hadamard gate is applied to all qubits except the first one
\begin{eqnarray}
        \ket{\phi}_{2}
        =\frac{1}{2^{NM}} \sum_{i=0}^{2NM-1}\ket{i}\otimes \ket{0}.\hspace{2.3cm}
\end{eqnarray}
We then apply $R_{y}\left(2\pi s_{1}^{T}x_{1}\right)$ on the last qubit when the remaining qubits are in the state
$\ket{0\cdots 00}_{23\cdots n}$ results in
\begin{eqnarray}
     \ket{\phi}_{3}= \frac{1}{2^{NM}} \sum_{i=0}^{2NM-1}\left(\cos\left(2\pi s_{1}^{T}x_{1}\right)\ket{2^{2NM}-1}\right.\nonumber\\\left.+\sin\left(2\pi s_{1}^{T}x_{1}\right)\ket{0} + \left( \sum_{i=1}^{2NM-1}\ket{i}\right)\right). \hspace{0.5cm}
\end{eqnarray}
Similar fashion, we continue apply multicontrolled unitary on the last qubit condition on all the possible combination of the remaing qubits with angle  $2\pi s_{2}^{T}x_{1},\cdots 2\pi s_{M}^{T}x_{1},2\pi s_{1}^{T}x_{2} \cdots 2\pi s_{M}^{T}x_{N}$, we obtain the desired result as represented in equation  \eqref{Encoded state}. If we express the amplitude of the encoded quantum state into $x_{j}^{m}$ and split the qubits into two index registers $\ket{j}$ and $\ket{m}$, then we can express our encoded quantum state in the following form
\begin{equation}
    \ket{\psi_{\mathbf{X}}}=\sum_{m=0}^{2M-1}\sum_{j=0}^{N-1}x_{j}^{m}\ket{m}\ket{j},
\end{equation}
Here, $x_{j}^{m}$ represents the value of the classical data at position $\left(j, m\right)$ in the data matrix $X$.
\begin{figure}
\includegraphics[width = 0.5 \textwidth]{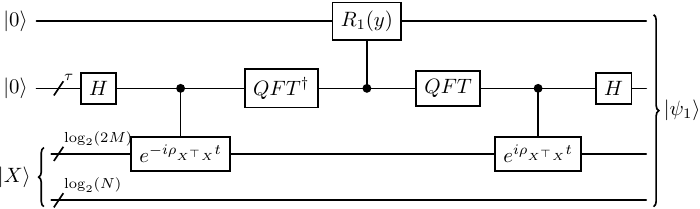}
\caption{\label{fig:2} Quantum principal component analysis (qPCA) is initially applied on $\rho_{\mathbf{X}^{\top}\mathbf{X}}$ to determine its eigenvalues and eigenvectors. Subsequently, a conditional rotation unitary is employed on the $\tau$ qubit register. Following this operation, the eigenvalue register and ancilla qubits are discarded, resulting in the state  $ \ket{\psi_{1}}=\frac{c_{1}}{\sqrt{p(1)}}\sum_{r=1}^{R}\frac{\lambda_{r}}{\lambda_{r}^{2}+\sigma_{n}^{2}} \ket{u_{r}} \ket{v_{r}}$. }
\end{figure}

\subsection{Eigenvalue and eigenstate estimation}
  Using Gram-Schmidt decomposition, we can re-expressed $\ket{\psi_{\mathbf{X}}}$ as \cite{WAN:17:PRA}
\begin{equation}
    \ket{\psi_{\mathbf{X}}}=\sum_{r=1}^{R}\lambda_{r}\ket{u_{r}}\ket{v_{r}},
    \end{equation}
We consider the density matrix $\rho_{\mathbf{X}^{\dagger}\mathbf{X}}=\operatorname{Tr}_{j}{\ket{\psi_{\mathbf{X}}}\bra{\psi_{\mathbf{X}}}}$ by disregarding the $\ket{j}$ register where $\operatorname{Tr}_{j}$ is the partial trace on $j$ qubits, which can be written as
\begin{equation}
    \rho_{\mathbf{X}^{\dagger}\mathbf{X}}=\operatorname{Tr}_{j}\{\ket{\psi_{\mathbf{X}}}\bra{\psi_{\mathbf{X}}}\}=\sum_{r=1}^{R}\lambda_{r}^{2}\ket{v_{r}}\bra{v_{r}}, 
\end{equation}
 Next, we apply the ideas of quantum principal component analysis  $\rho_{\mathbf{X}^{\dagger}\mathbf{X}}$ to $\ket{\psi_{\mathbf{X}}}$, resulting in the following expression \cite{LMR:14:NPh}
\begin{equation}
    \ket{\zeta_{1}}\bra{\zeta_{1}}=\sum_{k=0}^{K}\ket{k\Delta t}\bra{k \Delta t}\otimes e^{-\dot{\iota} k \rho_{\mathbf{X}^{\dagger}\mathbf{X}} \Delta t}\ket{\psi_{\mathbf{X}}}\bra{\psi_{\mathbf{X}}} e^{\dot{\iota} k \rho_{\mathbf{X}^{\dagger}\mathbf{X}} \Delta t},
\end{equation}
  for some large $\mathbf{K}$ and $\ket{\zeta_{i}}$ is $i$th intermediate quantum state.  By utilizing the quantum phase estimation algorithm, we can further write \cite{SSP:16:PRA}
  \begin{equation}
\ket{\zeta_{2}}=\sum_{r=1}^{R}\lambda_{r}\ket{u_{r}} \ket{v_{r}} \ket{\lambda_{r}^{2}},
  \end{equation}
where $\lambda_{r}$ encoded in the $\tau$ qubits of an extra register \cite{WAN:17:PRA}. 

\subsection{Mean of Gaussian process regression}
We employ the conditional unitary on the ancilla qubit to invert the singular values before estimating the mean of Gaussian process regression. This qubit is conditionally rotated based on the eigenvalue register
 \begin{eqnarray}
\ket{\zeta_{3}}=\sum_{r=1}^{R}\lambda_{r}\ket{u_{r}} \ket{v_{r}} \ket{\lambda_{r}^{2}} \left[\sqrt{1-\left(\frac{c_{1}}{\lambda_{r}^{2}+\sigma_{n}^{2}}\right)^2} \ket{0}\right. \nonumber \\ \left.+\frac{c_{1}}{\lambda_{r}^{2}+\sigma_{n}^{2}}\ket{1}\right],
   \end{eqnarray}
where the parameter $c_{1}$ is chosen such that the inverse eigenvalues remain bounded by 1. Subsequently, a conditional measurement is performed on the ancillary qubit. The algorithm proceeds only if the ancilla is measured in state $\ket{1}$. Discarding the eigenvalue register after measuring results in the state
 \begin{equation}
      \ket{\psi_{1}}=\frac{c_{1}}{\sqrt{p(1)}}\sum_{r=1}^{R}\frac{ \lambda_{r}}{\lambda_{r}^{2}+\sigma_{n}^{2}} \ket{u_{r}} \ket{v_{r}},
    \end{equation}
where the probability of acceptance is indicated by $p(1)=c_{1}\sum_{r}\left|\frac{ \lambda_{r}}{\lambda_{r}^{2}+\sigma_{n}^{2}}\right|^{2}$. The gate implementation, which outlines the process of transforming the quantum state from the singular value decomposition problem to the computation of eigenvalues and eigenvectors in the quantum register, is shown in Fig.~\ref{fig:2}.  

\begin{figure}
\includegraphics[width = 0.4 \textwidth]{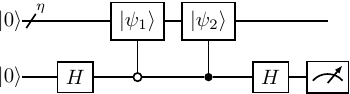}
\caption{\label{fig:3} Hadamard subroutine to estimate the mean of Gaussian process regression }
\end{figure}
We prepare another quantum state $\ket{\psi_{2}}=\ket{\phi_{*}}\ket{y}$ and use the Hadamard test between $\ket{\psi_{1}}$ and $\ket{\psi_{2}}$ to evaluate the mean as defined in Eq.~(\ref{Mean})
\begin{equation}
E\left[f_*\right]=c_{1}\sum_{r=1}^{R}\frac{\lambda_r}{\lambda_r^2 +\sigma^{2}}\braket{X_{*}|v_r} \braket{y|u_{r}}.
\end{equation}
 We obtain an expression equivalent to the Gaussian process regression mean up to some constant $c_{1}$. To obtain our desired result, we implement the Hadamard test subroutine, which is given in Fig.~\ref{fig:3}.  

\subsection{Variance of Gaussian Process regression}
For estimating the variance in Gaussian process regression, we use conditional rotation conditioned on the eigenvalue register such as

 \begin{eqnarray}
 &\ket{\zeta_{4}}   =\sum_{r=1}^{R} \ket{v_{r}}  \ket{u_{r}} \ket{\lambda_{r}^{2}} \nonumber\\ &\left[\sqrt{1-\left(\frac{c_2}{\lambda_{r}\sqrt{\lambda_{r}^{2}+\sigma_{n}^{2}}}\right)^2} \ket{0}+\frac{c_2}{\lambda_{r}\sqrt{\lambda_{r}^{2}+\sigma_{n}^{2}}}\ket{1}\right], 
 \end{eqnarray}

Performing measurement in ancilla register and proceed when the measurement in state $\ket{1}$. Discarding the quantum register result in the final state
     \begin{equation}
      \ket{\psi_{1}^{'}}=\frac{c_2}{\sqrt{p(2)} }\sum_{r=1}^{R}\frac{1}{\sqrt{\lambda_{r}^{2}+\sigma_{n}^{2}}}  \ket{v_{r}}  \ket{u_{r}},
    \end{equation}
where the probability of acceptance is given by $p(2)=c_2\sum_{r}\left|\frac{1}{\sqrt{\lambda_{r}^{2}+\sigma_{n}^{2}}}\right|^{2}$.

\begin{figure}
\includegraphics[width = 0.5 \textwidth]{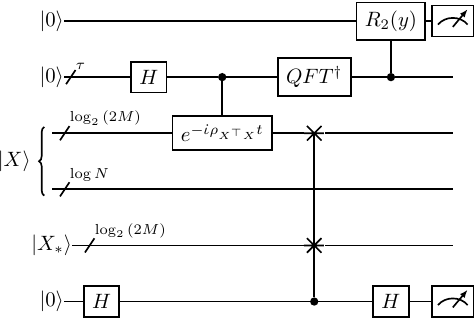}
\caption{\label{fig:4} We use qPCA along with QPE to extract the eigenvalues of the encoded data matrix, followed by the conditional rotation operators. We then perform the swap test between the singular vector of the data matrix and new data points and measure the first and last qubits. We proceed when the outcomes are in the $\ket{11}$ state. }
\end{figure}

Variance of the Gaussian process regression can only be positive, so we use the swap test to attain the desired outcome. Initially, we prepare the quantum state  $\ket{\psi_{2}^{'}}=\ket{\phi_{*}}$.  Employing a swap operation between  $\ket{\psi_{1}^{'}}$ and $\ket{\psi_{2}^{'}}$ we can calculate $\left|\braket{\psi_{1}^{'}|\psi_{2}^{'}}\right|^{2}$ which corresponds to the  posterior variance

\begin{equation}
   \mathrm{V}\left[f_*\right]= \frac{c_2^{2}}{p(2) }\sum_{r=1}^{R}\frac{1}{\lambda_{r}^{2}+\sigma_{n}^{2}}\left| \braket{v_{r}|\psi_{*}}\right|^{2}.
\end{equation}
We then multiply the noise variance $\sigma_{n}$ to obtain the variance of Gaussian process regression. Fig.~\ref{fig:4} shows the circuit implementation for the estimation of the variance of the Gaussian process regression.

\section{ \label{Sec:4} Results}
In this section, we analyze the computational complexity of our proposed method. The quantum state $\ket{\psi_{X}}$ can be prepared in computational complexity in $O\left(NM\right)$ in the first step. We then implement a quantum principal component analysis, which has a computational complexity of $O\left(\log (M) \epsilon^{-3}  \right)$ up to a desired error tolerance $\epsilon$. The next step of our algorithm, which is conditional unitary, can be implemented in $O \left(\log\left(\frac{1}{\epsilon}\right)\right)$, whose contribution is negligible.  
The computational complexity of the Hadamard test is $O\left(\frac{\log\left(1/\delta\right)}{\epsilon^{2}} \right)$. The total computational complexity of the algorithm then becomes $O\left(NM \log (M) \epsilon^{-5} \kappa^{2}  \log\left(1/\delta\right) \right)$.  

Our proposed model represents a significant advantage over existing classical and quantum methods. The computational complexity of classical methods bears a computational burden of $O\left(NM^{2}\right)$ \cite{SS:20:SC}.  Furthermore, recent advances in quantum-assisted Gaussian process regression have introduced a solution with time complexity $O\left[\kappa\left(\frac{1}{\sqrt{P_{k}}}DN\log\frac{1}{\delta}\log N \epsilon^{-3}+poly\log N\right)\right]$ where $P_{k}$ denotes the probability of success, $D$ is the dimension of the data point, and $\delta$ the precision of state preparation \cite{CYGYLGL:22:PRA}. Our quantum approach also outperforms this quantum solution.  

We also provide numerical simulations for our method, an open-source implementation of our simulation is available on GitHub\footnote{Source code available at: \href{https://github.com/cagalvisf/Quantum-RFF}
	{\url{https://github.com/cagalvisf/Quantum-RFF}}}.

We use a similar setting for quantum circuit simulation as described in \cite{FGS:24:PRA}, implementing the quantum-assisted Hilbert space (QHS) Gaussian process regression. The accuracy of the algorithm is sensitive to the time parameter $t=2\pi/\delta_{R}$, where we select $\delta_{R}$ slightly larger than the maximum eigenvector $\lambda_{max}$ of the operator $\rho_{\mathbf{X}^{\top}\mathbf{X}}$. We use the constant parameter $c_{1}=\lambda_{max}^2+\sigma^{2}$ and $c_{2} = \lambda_{max}\sqrt{\lambda_{max}^{2}+\sigma^{2}}$ in our simulation.

Our experiment used $N=16$ data measurements with random Gaussian noise parameter $\sigma_n = 0.1$ to approximate the function $f(x) = \sin(x)$. The Gaussian process regression Kernel has hyperparameters length scale $l=1$ and signal variance $\sigma_{f}=1.5$. We select
$10^{6}$ shots and $\tau=13$ qubits for our eigenvalue estimation register. 
\begin{figure}
    \centering
    \includegraphics[width=0.9\linewidth]{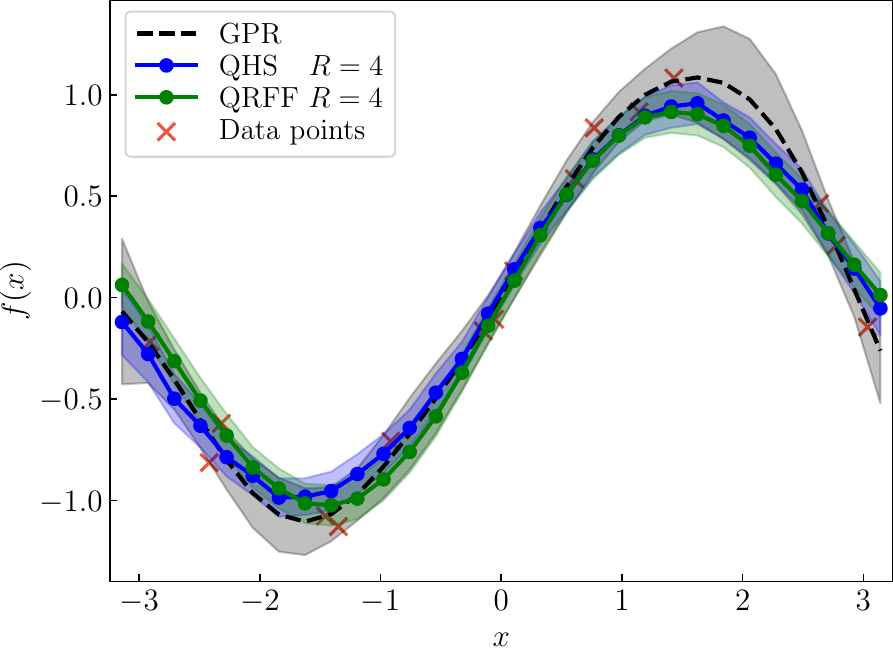}
    \caption{Mean and variance of the Gaussian process regressions given the $N=16$ data points. The black dashed curve shows the traditional GPR method. The blue line corresponds to the QHS regression method in \cite{FGS:24:PRA} and the green line is the quantum random Fourier feature (QRFF) method proposed in this paper.}
    \label{fig: mean and variance}
\end{figure}

The simulation plots in Fig. \ref{fig: mean and variance} compare the quantum RFF method with the classical GPR and the already mentioned QHS method. The quantum-assisted random Fourier feature-based Gaussian process regression mean exhibits similar behavior to the classical method, thereby verifying the effectiveness of our algorithm. However, the variance of the approximations is distant from the expected variance. This limitation is also evident in the classical RFF and HS approximation methods for GPR. To address this issue, it is essential to increase the approximation parameter $M$. However, the constraints of simulating qubits on a classical computer require the use of a small approximation parameter.

\section{\label{Sec:5} Conclusion}
We have proposed a novel hybrid approach for reduced-rank quantum-assisted Gaussian process regression. We efficiently approximate the kernel function on a classical computer using a random Fourier feature method. We use multi-controlled rotation to load classical data into a quantum computer efficiently. Our method utilizes quantum principal component analysis to extract eigenvectors and eigenvalues. A controlled rotation operation is performed based on the eigenvalue register. We then employ the Hadamard and Swap tests to estimate the mean and variance values of the Gaussian process regression.  

The deterministic nature of our algorithm for quantum state preparation makes it more suitable for modeling real-world applications using Gaussian process regression. Our algorithm is susceptible to the time parameter $t$ in simulations. A careful selection of $t$ is important to run the model successfully. We can design a strategy to optimize the $t$ parameter for robust solutions in future work.

\section*{Acknowledgment}
C.A. Galvis-Florez, A. Farooq, and Simo Särkkä thank the Research Council of Finland for funding this research (project 350221). C.A. Galvis-Florez and A. Farooq contributed equally to this work. All the authors contributed to the original idea, writing, and finalizing the paper results.

\balance
\bibliographystyle{ieeetr}
\bibliography{Bibliography}

\end{document}